\documentclass[showpacs,preprint,pre]{revtex4}
\usepackage{graphicx}
\begin{document}

\begin{abstract}
A 1-dimensional model for coherent quantum energy transfer through
a complex of compressible boxes is investigated by numerical integration
of the time-dependent Schr\"odinger equation.  Energy is communicated
from one box to the next by the resonant fluctuating Fermi pressure of the
electrons in each box pushing on the walls and doing work on adjacent
boxes.  Parameters are chosen
similar to the chain molecules of typical light harvesting
complexes.  For some parameter choices the system is found to have
an instability leading to self-induced coherent energy transfer transparency.
\end{abstract}

\title{Model for energy transfer by coherent Fermi pressure fluctuations in quantum soft matter}
\author{Mark A. Peterson}
\affiliation{Mount Holyoke College}
\email{mpeterso@mtholyoke.edu}
\pacs{87.15.ag, 87.15.hg, 82.20.Ln, 03.65.Aa}
\date{\today}

\maketitle
\section{Introduction}
Light harvesting complexes in plants and bacteria, as well as visual systems in
animals, are typically arrays of chromophores organized spatially and
held in place by proteins (see recent review articles \cite{Scholes},\cite{Rhodopsins}
and references there).  They
absorb photons and transfer the excitation energy over distances as great as
several nanometers with high quantum efficiency, estimated at 67\% in the
mammalian visual system, and up to 90\% in the photosynthesis of higher plants.
No artificial solar cell approaches these values.  Biological systems
achieve these values not just through being nanoscale, but also through being ultrafast,
much faster than corresponding radiative lifetimes and other dissipative
mechanisms.
The isomerization
of retinal in rhodopsin, the first step in vision, is one of the
fastest photochemical reactions known, complete in 200 fs.  A similar sub-picosecond
timescale characterizes the first energy transfers in photosynthesis.
Understanding these fast energy pathways is a subject of intensive current research.

In principle the chain of events following photoabsorption could be just incoherent transitions
from one excited state to another, passing energy to a reaction center (RC) in photosynthesis,
or to a new conformation of rhodopsin in vision.  The speed of these processes, however, together
with the experimental observation of coherences among some intermediate states, has prompted
the investigation of coherent, quantum mechanical evolution of the excited states as a
contributing mechanism.  These investigations have even raised fundamental issues
of quantum mechanics, suggesting that coherent energy transfer could be a kind of quantum
computer, finding the best path by, in effect, trying all of them
simultaneously \cite{Engel2007}, or questioning
whether coherent mechanisms seen in the laboratory have any relevance to performance under
incoherent illumination by sunlight \cite{Fassioli2012}.
The decay of coherence through interaction with the
environment is relevant to such questions, and has been modeled by Hamiltonians that
include interaction of the electronic state with `phonons,' vibrational modes of the molecules,
reviewed in \cite{Fleming2009}.

This paper calls attention to an energy transfer mechanism that has not been part of
the discussion.  The excitonic state of a chromophore
itself has a kind of phonon-like quality, even in the absence of nuclear motion,
manifesting itself as an oscillating Fermi pressure.  By steric interaction with a
neighboring molecule, this oscillating pressure can do work and transfer energy.
An essential feature of the model is that each molecule sees each neighboring molecule
as a wall, in the sense of classical mechanics,
into which it cannot penetrate. (The reason for this is the Pauli principle, but the
quantum mechanical origin of this excluded space is otherwise irrelevant to the
molecular quantum states.)  The motion of the wall, being a collective motion of the
molecular electrons, will be essentially classical. In the model that we compute,
the walls, considered as classical oscillators, have a natural frequency
much lower than the exciton resonance frequency, another reason that their motion
should be considered classical.  The oscillating pressure, on the other hand,
is entirely of quantum mechanical origin, as the beating of the molecular excited
state with the ground state in a coherent superposition.

Biological systems do not suggest any particular geometry for light harvesting antennae beyond
the close proximity of chromophores and proteins.  Accordingly we choose the simplest possible
geometry to test the concept, in what is surely
the simplest possible model of electronic excitation coupled to steric
interaction in this way.
It is exactly solvable in the time domain by numerical integration of the
Schr\"odinger equation.
The model is not proposed as a realistic geometry for an actual process, but only as a
schematic way of thinking about a process that undoubtedly does occur at some level.
We are not doing molecular modeling, but rather a much more primitive thing,
investigating whether the
proposed mechanism could be significant under any circumstances, and might therefore
be worth detailed consideration in more realistic geometries.

The model is a chain of 1-dimensional boxes, each box representing a single chromophore.
The relevant electronic states are quantum mechanical box states, where we allow the
box to fluctuate in size, subject to the Fermi pressure of the electrons.
The walls of the boxes represent the boundaries between chromophores, with masses
assumed to be tens or hundreds of electron masses.
In this way one box can push on the next, compressing it and raising its energy levels,
or perhaps inducing quantum transitions.  The mechanism is non-Coulombic, and universal
in the sense that it is not constrained by selection rules.  In particular it would work
the same way for both optically allowed and optically forbidden states, if we included
enough molecular detail to make that distinction.

A computation shows that energy can be tranferred from box to box by this process, and in
an unexpected way.  As we show by computation, there is
an instability in this system that allows the growth of fluctuations in Fermi pressure
to feed on itself and grow to dominate the time development, creating a self-induced
energy transfer transparency through the system.  For comparison
we include also energy transfer by the familiar F\"orster resonance mechanism, and find
that, within this model,
the Fermi pressure mechanism can be of comparable importance.

\section{Particle in a compressible box}
Consider a quantum mechanical particle of mass $\mu$ in one dimension in a box that occupies
the time-dependent interval $[L(t),R(t)]$ on the x-axis.  This very elementary system has been treated
before %
\cite{Tarr},
but we describe it again here for
completeness.  The wave function
$\Phi(x,t)$ satisfies the time-dependent Schr\"odinger equation
\begin{equation}
\label{SE1}
i\hbar\frac{\partial\Phi}{\partial t}=-\frac{\hbar^2}{2\mu}\frac{\partial^2\Phi}{\partial x^2},
\end{equation}
in the box $[L,R]$, as well as the boundary conditions
\begin{equation}
\Phi(L(t),t)=\Phi(R(t),t)=0.
\end{equation}
In terms of the variable
\begin{equation}
\xi(x,t)=\frac{x-L}{R-L}
\end{equation}
we can seek a solution to Eq.~(\ref{SE1}) in the form
\begin{equation}
\Phi(x,t)=\Psi(\xi,t)
\end{equation}
where $\Psi$ satisfies
\begin{equation}
\label{SE2}
i\hbar\left[\frac{\partial\Psi}{\partial t}-\frac{1}{R-L}\frac{\partial\Psi}{\partial\xi} [\xi(\dot{R}-\dot{L})+\dot{L}]\right]=-\frac{1}{(R-L)^2}\frac{\hbar^2}{2\mu}
\frac{\partial^2\Psi}{\partial\xi^2}
\end{equation}
in the interval $0<\xi<1$, and the boundary conditions
\begin{equation}
\label{BC2}
\Psi(0,t)=\Psi(1,t)=0
\end{equation}

The wavefunction $\Phi$ determines the probability density $|\Phi|^2$.
In terms of the new variable $\xi$ the corresponding probability density
is $|\Psi|^2(R-L)$.  We verify unitarity in the form
\begin{eqnarray}
0&=&\frac{d}{dt}\int_0^1\Psi^*\Psi(R-L)d\xi\\
&=&\int_0^1\left[
\frac{\partial\Psi^*}{\partial t}\Psi+\Psi^*\frac{\partial\Psi}{\partial t}
+\Psi^*\Psi\frac{\dot{R}-\dot{L}}{R-L}
\right](R-L)d\xi
\end{eqnarray}
for all $t$, using Eq.~(\ref{SE2}) and several integrations by parts.

For each $t$ the wavefunction $\Psi$ belongs to the Hilbert space
of square integrable functions on the interval $[0,1]$, vanishing at the endpoints.
The natural inner product $<~|~>$ on the Hilbert space at $t$ is
\begin{equation}
<f|g>=(R-L)\int_0^1 f^*(\xi)g(\xi)d\xi
\end{equation}
and the functions
\begin{equation}
\Psi_m(\xi,t)=\frac{\sqrt{2}\sin m\pi\xi}{\sqrt{R-L}}\quad\quad m=1,2,3,...
\end{equation}
are an orthonormal basis at each $t$.

Expand the wavefunction $\Psi$ in this basis
\begin{equation}
\Psi=\sum_m a_m(t)\Psi_m.
\end{equation}
Inserting this representation into Eq.~(\ref{SE2}) and resolving
the terms in the basis $\{\Psi_m\}$ leads to a representation of
the Schr\"odinger equation as a system of ordinary differential
equations for the amplitudes $a_m$,
\begin{equation}
\label{SE3}
\frac{d a_m}{dt} =\frac{2}{R-L}\sum_{n\neq m}\frac{mn}{m^2-n^2}[\dot{L}-\dot{R}(-1)^{n+m}]a_n - \frac{i\hbar m^2 \pi^2}{(R-L)^2 2\mu}a_m
\end{equation}
Equivalently one can write this system in the ``interaction representation," defining
\begin{eqnarray}
b_m&=&a_m e^{im^2\phi(t)}\\
\label{dphi}
\phi(t)&=&\frac{\hbar\pi^2}{2\mu}\int_0^t\frac{1}{(R-L)^2}\,dt
\end{eqnarray}
Eq.~(\ref{SE3}) becomes
\begin{equation}
\label{SE3hat}
\frac{d b_m}{dt} =\frac{2}{R-L}\sum_{n\neq m}e^{i(m^2-n^2)\phi}\frac{mn}{m^2-n^2}[\dot{L}-\dot{R}(-1)^{n+m}]b_n
\end{equation}
In terms of these amplitudes the statement of unitarity takes the form
\begin{equation}
\frac{d}{dt}\sum_m a_m^*a_m=0,
\end{equation}
easily verified using Eq.~(\ref{SE3}), and similarly for the analogous statement in
terms of $b_m$.

The expectation value of the energy
\begin{equation}
<E>=\sum_m a_m^*a_m \frac{\hbar^2 m^2 \pi^2}{2\mu(R-L)^2}
\end{equation}
is time dependent because the moving walls can do work on the particle.
In fact, using Eq.~(\ref{SE3}), we find
\begin{eqnarray}
\frac{d<E>}{dt}&=&\frac{\hbar^2\pi^2}{\mu(R-L)^3}\sum_{m,n}mn[\dot{L}-\dot{R}(-1)^{n+m}]\Re(a_n^*a_m)\\
  &=&\frac{\hbar^2\pi^2}{\mu(R-L)^3}\left(\dot{L}\left|\sum_n{na_n}\right|^2-\dot{R}\left|\sum_n{(-1)^nna_n}\right|^2\right)
\end{eqnarray}
The terms with $n \neq m$ are rapidly fluctuating, while the terms with $n=m$ simply
reflect the adiabatic compression or expansion of the box.
The coefficients of $\dot{L}$ and $-\dot{R}$ can be interpreted as the pressures
at the left and right walls respectively.  It is exactly the fluctuating component of these
pressures that will be responsible for the energy transfer in the model that we compute.

\section{2N electrons in a box with moving walls}
\label{2Nelectrons}
The time development of multielectron states can be described by
regarding Eqs.~(\ref{SE3}) and (\ref{SE3hat}) as equations for the
evolution of fermionic operators (we develop this point in more detail in
section \ref{op}).  The ground state of a box with
$2N$ electrons would then be
\begin{equation}
|N\rangle = a^\dagger_{N\uparrow}a^\dagger_{N\downarrow}...a^\dagger_{1\uparrow}a^\dagger_{1\downarrow}|0\rangle
\end{equation}
and one of the (doubly degenerate) first excited states would be
\begin{equation}
|N^*\rangle = a^\dagger_{N+1\uparrow}a^\dagger_{N\downarrow}...a^\dagger_{1\uparrow}a^\dagger_{1\downarrow}|0\rangle,
\end{equation}
where $|0\rangle$ is the empty box, and where the electron spin ($\uparrow$
and $\downarrow$)
plays no essential role in what follows and will be
largely ignored.  Restricting attention to just the transition between
these two states (HOMO-LUMO transition), and suppressing the subscript for spin, we
have the time-dependent Schr\"odinger equation in the form
\begin{eqnarray}
\frac{d a_N}{dt} &=&\frac{-2}{R-L}\frac{N(N+1)}{2N+1}(\dot{L}+\dot{R})a_{N+1}
 - \frac{i\hbar N^2 \pi^2}{(R-L)^2 2\mu}a_N\\
\frac{d a_{N+1}}{dt} &=&\frac{2}{R-L}\frac{N(N+1)}{2N+1}(\dot{L}+\dot{R})a_N
 - \frac{i\hbar (N+1)^2 \pi^2}{(R-L)^2 2\mu}a_{N+1}
\end{eqnarray}
or equivalently, in the interaction picture,
\begin{eqnarray}
\frac{d b_N}{dt} &=&\frac{-2}{R-L}\frac{N(N+1)}{2N+1}e^{-i(2N+1)\phi}(\dot{L}+\dot{R})b_{N+1}\\
\frac{d b_{N+1}}{dt} &=&\frac{2}{R-L}\frac{N(N+1)}{2N+1}e^{i(2N+1)\phi}(\dot{L}+\dot{R})b_N
\end{eqnarray}
The time evolution of $\phi$ is still given by Eq.~(\ref{dphi}).
If the box is squeezed symmetrically ($\dot{L}=-\dot{R}$), no transition
is excited, but the energy goes up.  It is amusing to note that if the
box is shaken rigidly ($\dot{L}=\dot{R}$) at the frequency
\begin{equation}
\omega_{res}=(2N+1)\dot{\phi}=\frac{(2N+1)\hbar\pi^2}{2\mu(R-L)^2}
\end{equation}
Rabi oscillations are excited, i.e., oscillations in the amplitudes
of the two-level system at a frequency proportional
to the amplitude of the resonant driving force.

The expectation value of the energy of the system is
\begin{equation}
\langle E\rangle=\frac{\hbar^2}{2\mu(R-L)^2}
\left[\frac{N(2N^2+1)}{3}+N^2|a_N|^2+(N+1)^2|a_{N+1}|^2\right]
\end{equation}
and its rate of change due to the work done by the walls is
\begin{equation}
\frac{d\langle E\rangle}{dt}=P_L\dot{L}-P_R\dot{R}
\end{equation}
where the ``pressures" are
\begin{eqnarray}
\label{PL}
P_L&=&\frac{\hbar^2\pi^2}{\mu(R-L)^3}\left[\frac{N(2N^2+1)}{3}+\left|(N+1)a_{N+1}+Na_N\right|^2\right]\\
\label{PR}
P_R&=&\frac{\hbar^2\pi^2}{\mu(R-L)^3}\left[\frac{N(2N^2+1)}{3}+\left|(N+1)a_{N+1}-Na_N\right|^2\right]
\end{eqnarray}

Note that the fluctuating component of pressure is proportional to $|a_Na_{N+1}|$, so that it is
non-zero only in a superposition of excited and ground states.

\section{2N electrons in each of J boxes}
\label{JBoxes}
If there are J boxes, the Hilbert space of states is the J-fold
tensor product of the Hilbert space for a single box.  The dynamics
is just that of the
formalism already described, but with
subscripts on all quantities labeling which box they belong to.
We assume no tunneling
between boxes (i.e., the wave functions vanish at
L and R as before).  The dynamics takes place independently in each
box, and does not, for example, lead to entanglement of the states,
but the systems may be coupled through the motions
of the walls.  In particular, the case we shall consider, the boxes may
be concatenated together, so that $0=L_1<R_1=L_2=x_1<R_2=L_3=x_2< ... <R_{J-1}=L_J=x_{J-1}<R_J=x_J=J$.
Thus the $J$ boxes occupy the interval $[0,J]$ on the x-axis, and
their average length is 1.

We also enlarge the dynamical system to include the moveable walls at
$x_1,x_2,...,x_{J-1}$, while keeping $x_0=0$ and $x_J=J$ fixed.
We treat the moveable walls classically,
giving them masses $M_1, M_2,...,M_{J-1}$ sufficiently large, and
imagine that the relevant forces on them are just the pressures
due to the delocalized electrons of the previous section, i.e.
\begin{eqnarray}
\label{dxdt}
\frac{dx_j}{dt}&=&v_j\\
\label{dvdt}
\frac{dv_j}{dt}&=&(P_{R_j}-P_{L_{j+1}})/M_j
\end{eqnarray}
There is a conserved energy in this system,
\begin{equation}
E=\sum_{j=1}^J{\langle E_j\rangle} + \frac{1}{2}\sum_{j=1}^{J-1} M_j v_j^2,
\end{equation}
useful for checking correctness of numerical computations.

In light harvesting complexes the resonant frequencies (the energy scale) of the
observable optical transitions typically decrease
in the direction that the excitation energy follows to the RC.
It was once suggested that this amounted to a kind of ``energy funnel," but the funnel is
not unidirectional, according to
more recent ideas:  the photosynthetic complex may be more like a reservoir
in which the captured energy is distributed \cite{Ritz}.  We can build such a structure into
the chain of J boxes by choosing effective electron masses $\mu_1<\mu_2< ...< \mu_J$
that increase slightly in the direction that we expect the energy to flow.
This alters the energy levels so that they are not initially in resonance, creating the
situation that we had set out to investigate.

Imagine that all boxes are in their quantum mechanical ground state.
Mechanical equilibrium in Eqs.~(\ref{dxdt})-(\ref{dvdt}) requires
\begin{equation}
\frac{x_j-x_{j-1}}{x_{j-1}-x_{j-2}}=\left(\frac{\mu_{j-1}}{\mu_j}\right)^{1/3}, \quad\quad j=2,..,J.
\end{equation}
Thus if the $\mu_j$'s increase with $j$, the boxes also become more compressed
with $j$, in order to balance the pressure at each wall.
Their transitions are not in resonance, however, because resonance between box $j$ and box $j-1$
requires
\begin{equation}
\frac{x_j-x_{j-1}}{x_{j-1}-x_{j-2}}=\left(\frac{\mu_{j-1}}{\mu_j}\right)^{1/2},
\end{equation}
with box $j$ slightly more compressed than it is in mechanical equilibrium.

The following scenario motivated the computation.
Taking the equilibrium state as the starting point, imagine that box 1 is suddenly
placed into its excited state by absorption of a photon.  The pressure is now higher
in box 1, so that box 2 begins to be compressed.  At sufficient compression box 2 comes into resonance with box 1,
and its excited state is populated by the oscillating pressure
through a Rabi transition.  As the pressure in box 2 increases,
box 3 begins to be compressed, etc., passing the excitation along the chain.  Computation shows
that something like this happens, but also that the classical intuition is not completely correct.
Rather, low frequency oscillations of the walls, induced by the initial photoabsorption,
bring the boxes in and out of resonance.
In the
assumed pure starting states the pressure in each box is nearly constant, and transitions are driven
only weakly, even during the resonances.  As the states gradually become
coherent superpositions of excited and ground states, an instability is reached in which
the rapidly oscillating pressure, growing in amplitude, takes over the time development, and energy is then rapidly transferred
through the whole complex.

\section{Operator formalism and the F\"orster term}
\label{op}
This section elaborates on the operator formalism suggested in section \ref{2Nelectrons}.
The Fermionic operators $a_m$ and their adjoints obey canonical anticommutation relations
\begin{equation}
\{a^\dagger_m,a_n\}=\delta_{mn},\quad\quad\{a_m,a_n\}=\{a^\dagger_m,a^\dagger_n\}=0,
\end{equation}
the empty box state obeys
\begin{equation}
a_m |0\rangle=0,
\end{equation}
(and similarly for the operators $b_m$ of the interaction picture) for all $m$.
A second subscript will indicate in which box the operator operates, as in $a^\dagger_{m,j}$, the creation operator
for the $m$th state in the $j$th box.
Operators for different boxes commute, since they operate on different factors in the
tensor product.

Hamiltonian operators can be written in these terms.  The kinetic energy operator is
\begin{eqnarray}
H_0&=&\sum_{spins}\sum_{j=1}^J\sum_{m=1}^{N+1}E_{m,j} a^\dagger_{m,j} a_{m,j}\\
E_{m,j}&=&\frac{\hbar^2\pi^2m^2}{2\mu_j(R_j-L_{j})^2},
\end{eqnarray}
The effect of the moving walls
on the time development can be given a Hamiltonian form
\begin{eqnarray}
H_1&=&i\hbar\sum_{j=1}^J\frac{2 N(N+1)}{(R_j-L_{j})(2N+1)}(\dot{L}_{j}+\dot{R_j})
\left(a^\dagger_{N+1,j}a_{N,j}-a^\dagger_{N,j}a_{N+1,j}\right)\\
&=&i\hbar\sum_{j=1}^J\left(\alpha_j b^\dagger_{N+1,j}b_{N,j}-\alpha_j^*b^\dagger_{N,j}b_{N+1,j}\right),{\rm ~where}\\
\alpha_j&=&\frac{2}{R_j-L_{j}}\frac{N(N+1)}{2N+1}(\dot{L}_{j}+\dot{R}_j)e^{i(2N+1)\phi_j}.
\end{eqnarray}
Finally we can include the F\"orster Hamiltonian, which couples nearest neighbor dipole moments,
\begin{equation}
H_{dipole-dipole}=\frac{\vec{\mu}_j\cdot\vec{\mu}_{j+1}}{|r_j-r_{j+1}|^3}
\end{equation}
The relevant part of the dipole moment of the $j$th box is
\begin{equation}
e\langle N^*| \vec{{\rm x}} |N \rangle_j=e(R_j-L_j)2\int_0^1\xi \sin(N\pi\xi)\sin((N+1)\pi\xi)\,d\xi=-e(R_j-L_j)\frac{8N(N+1)}{\pi^2(2N+1)^2}.
\end{equation}
Thus the F\"orster Hamiltonian is
\begin{eqnarray}
\label{H2}
H_2&=&i\hbar\sum_{j=1}^{J-1}\left[\beta_j b^\dagger_{N+1,j+1}b_{N,j+1}\otimes b^\dagger_{N,j}b_{N+1,j}
-\beta^*_j b^\dagger_{N,j+1}b_{N+1,j+1}\otimes b^\dagger_{N+1,j}b_{N,j}\right], {\rm~where}\\
\label{beta}
\beta_j&=&e^2\gamma_j\left[\frac{8N(N+1)}{\pi^2(2N+1)^2}\right]^2(R_{j+1}-L_{j+1})(R_j-L_j)\frac{2}{(R_{j+1}-L_j)^3} e^{i(2N+1)(\phi_{j+1}-\phi_{j})}
\end{eqnarray}
and $\gamma_j$ is a real phenomenological factor of order 1 which could have
either sign, depending on the mutual orientation of the dipoles.

\section{Choices of Parameters}
We follow Ref.\cite{Phillips}
in modeling a typical light harvesting molecule
as a box built out of $2N=28$ units (i.e., $N=14$), each unit contributing 1 electron to the delocalized states, and
each of length $a=0.1$~nm.  The resonant frequency is then
\begin{equation}
\omega_{res}=\frac{\hbar\pi^2(2N+1)}{2\mu[(2N-1)a]^2}=2.3\times 10^{15} {\rm~s}^{-1}
\end{equation}
corresponding to a wavelength $\lambda=830$~nm.
Choose units in which $\hbar=m_e=(2N-1)a=1$.  The unit of time is then
\begin{equation}
\frac{(2N-1)^2a^2m_e}{\hbar}=63~{\rm fs} \quad\quad{\rm if~N=14}
\end{equation}
We have made the conversion to
physical time in reporting the
course of the excitation through the chain of boxes.


F\"orster transfer is the incoherent transfer of excitation by the mechanism of $H_2$,
usually calculated in time-dependent perturbation theory by Fermi's golden rule.  Over
the short times that we are modeling we instead consider $H_2$ to contribute
to the coherent time development.  We can choose parameters so that
the coherent transfer time due to $H_2$ alone is a typical F\"orster transfer time, say 5 ps.  Let there be just 2 boxes,
rigid, each of length 1, so that they are in resonance.  $H_2$ then drives oscillations at the
frequency
\begin{equation}
\omega_2=e^2\gamma_1\left[\frac{8N(N+1)}{\pi^2(2N+1)^2}\right]^2\approx \frac{4e^2}{\pi^4},
\end{equation}
and this is also the coupling constant in $H_2$.  If we want the corresponding half period $\pi/\omega_2$ to
be 5 ps (physically), or $5/0.063\approx 79$ in our dimensionless units, then the coupling
constant must be
\begin{equation}
\label{e2}
\frac{4e^2}{\pi^4}\approx \pi/79 \approx 0.04.
\end{equation}

A common sense check of this parameter value comes from the virial theorem.  The total kinetic
energy of the 2N delocalized electrons is $N(N+1)(2N+1)\pi^2/6$, and therefore the total
electrostatic potential energy should be
\begin{equation}
\frac{N(N+1)(2N+1)\pi^2}{3}=(2Ne)(2Ne_{nuc})\left\langle\frac{1}{r}\right\rangle,
\end{equation}
If we use the coupling constant from Eq.~(\ref{e2}), and take the effective charge of the screened
nuclei in each unit to be $e_{nuc}=e$, we find, still using $N=14$,
\begin{equation}
\left\langle \frac{1}{r}\right\rangle\approx \frac{N(N+1)(2N+1)}{3N^2\pi^2(0.04)}\approx 26
\end{equation}
as if the chain molecule had length 1, by choice of units, but width only 1/2N, a reasonable picture.

\section{Algorithms and Results}
We integrate the Schr\"odinger equation
\begin{equation}
i\hbar\frac{d\Psi}{dt}=(H_1+H_2)\Psi
\end{equation}
in the interaction picture, described above by the operators $b_m$, arguing that
over the short times that we will investigate, $H_2$ should also contribute
to the coherent time development of the quantum state of the system.  The Hamiltonian $H_2$ entangles
the box states, so that it is no longer possible
to treat the dynamics in each box separately.  Where the state space could have dimension $2J$
in section \ref{JBoxes}, it now must have dimension $2^J$, a notable increase in complexity.  Let us
choose a basis consisting of tensor products of J box states, each
being one of $|N\rangle$ and $|N^*\rangle$, with the Jth box represented at the left and the 1st box at the right,
and let us label them by integers $0,1,...,2^J-1$.  The labels are read as follows:  express the label
in binary, and interpret the digits 1 or 0, left to right, as meaning the excited state or the ground state.
Thus the label $13=1101_2$ is for the state $|N^*\rangle\otimes|N^*\rangle\otimes|N\rangle\otimes|N^*\rangle$.

It is now straightforward to write out the matrix of $H_1+H_2$ in the interaction picture.  For $J=2$ it is
\begin{equation}
H_1+H_2=i\hbar\left(\begin{array}{cccc}
                      0 & -\alpha^*_1 & -\alpha^*_2 & 0\\
                      \alpha_1 & 0 & -\beta_1^* & -\alpha_2^* \\
                      \alpha_2 & \beta_1 & 0 & -\alpha_1^* \\
                      0 & \alpha_2 & \alpha_1 & 0
                    \end{array}\right)
\end{equation}
More generally one can prove the following inductive scheme for constructing the matrix of $H_1+H_2$
(apart from the factor $i\hbar$)
for any $J$.  Let $A_J$ be the $2^J\times 2^J$ matrix for $(H_1+H_2)/i\hbar$ in the case of J boxes.  Let $B_J$ be the
$2^J\times 2^J$ matrix given by (in Matlab notation)
\begin{equation}
B_J={\rm diag}(\alpha_{J+1}*{\rm ones}(1,2^J))+{\rm diag}(\beta_{J}*{\rm ones}(1,2^{J-1}),2^{J-1})
\end{equation}
i.e., $\alpha_{J+1}$ on the diagonal, and $\beta_{J}$ on a superdiagonal.  Then
\begin{equation}
A_{J+1}=\left(\begin{array}{cc}
                A_J & -B_J^\dagger \\
                B_J & A_J
              \end{array}
              \right)
\end{equation}

The pressures $P_{L_{j}}$ and $P_{R_j}$ in box $j$ are still given by Eqs.~\ref{PL}-\ref{PR}, using the
variables of box $j$, where it is now
understood that we must trace over all the other boxes $j'\neq j$.  An efficient way to compute these pressures
is to find the coefficients of $\dot{L}_{j}$ and $-\dot{R}_j$ in
\begin{equation}
\frac{d\langle E\rangle}{dt}=\left\langle \frac{dH_0}{dt}\right\rangle+\frac{1}{i\hbar}\langle\left[H_0,H_1\right] \rangle
\end{equation}
Thus, for example, if $J=2$ and the wavefunction is
\begin{equation}
\Psi=\sum_{m,n=0}^{1}c_{mn}\Psi_{mn},
\end{equation}
using the binary notation for the labels,
\begin{eqnarray}
P_{L_1}&=&\frac{\hbar^2\pi^2}{\mu_1(R_1-L_1)^3}\left[\frac{N(2N^2+1)}{3}+\left|Nc_{00}+(N+1)c_{01}\right |^2
 +\left |Nc_{10}+(N+1)c_{11}\right |^2 \right]\\
P_{R_1}&=&\frac{\hbar^2\pi^2}{\mu_1(R_1-L_1)^3}\left[\frac{N(2N^2+1)}{3}+\left|Nc_{00}-(N+1)c_{01}\right |^2
 +\left |Nc_{10}-(N+1)c_{11}\right |^2 \right]
\end{eqnarray}

We have investigated chains of 4 boxes, with slight systematic trends in their resonance frequencies
from one box to the next, parameterized by systematic trends in the effective electron masses
\begin{equation}
\mu_j=1+j\Delta \mu \quad\quad j=0,1,2,3
\end{equation}
where the detuning parameter $\Delta\mu$ takes values
$0.012\leq\Delta\mu\leq 0.04$.

\begin{figure}
\includegraphics{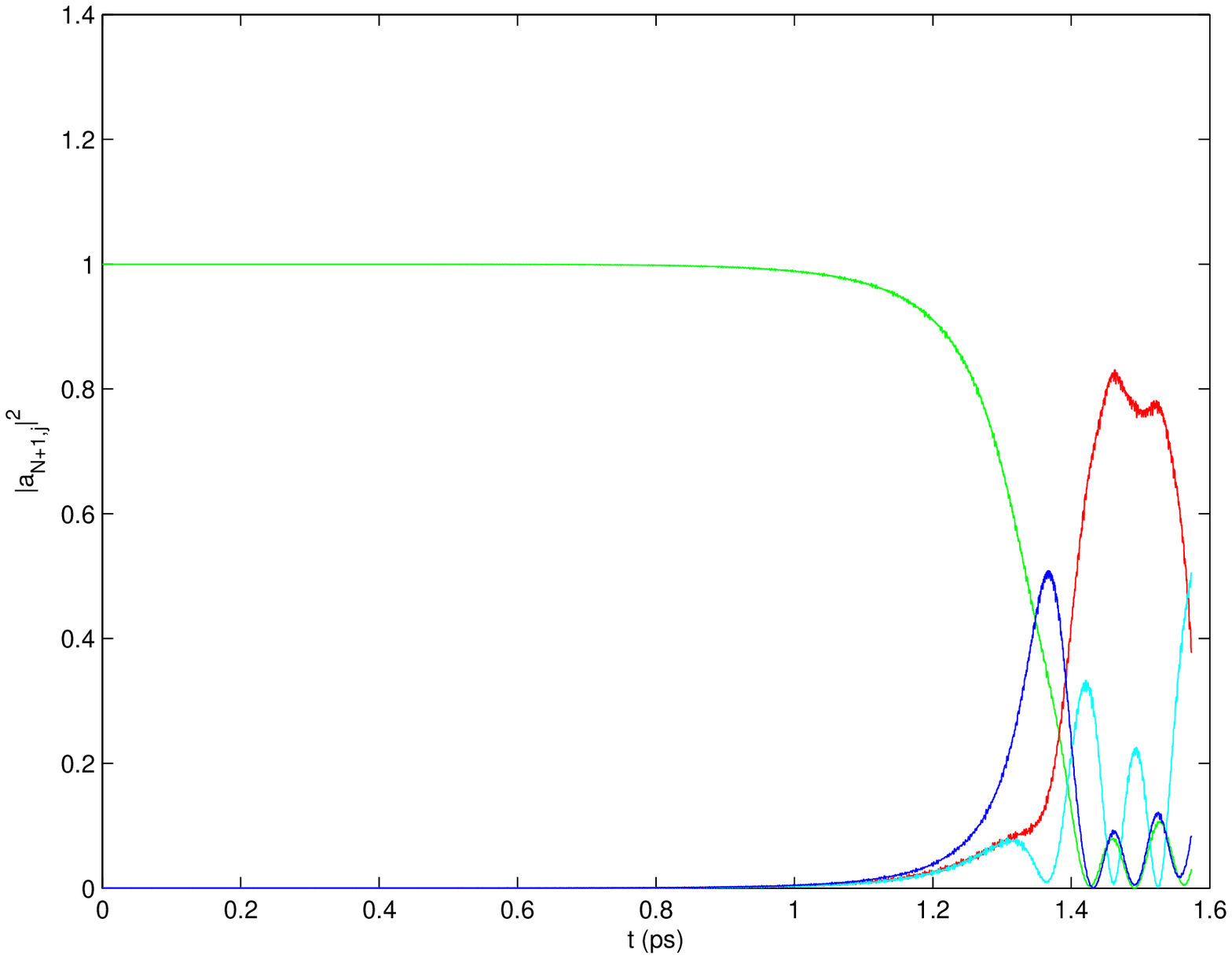}
\caption{The excitation $|a_{N+1}|^2$ as a function of time in box 1 (green), box 2 (red), box3 (cyan)
and box 4 (blue).  Here N=14, and the effective masses, creating the funnel,
were $\mu_1=1$, $\mu_2=1.02$, $\mu_3=1.04$, and $\mu_4=1.06$.  All wall masses were M=100.
The excitation reaches the last (4th) box directly from the first box,
with 50\% strength, at about 1.3 ps. The F\"orster term is not included.}
\label{gamma0Mass100}
\end{figure}

\begin{figure}
\includegraphics{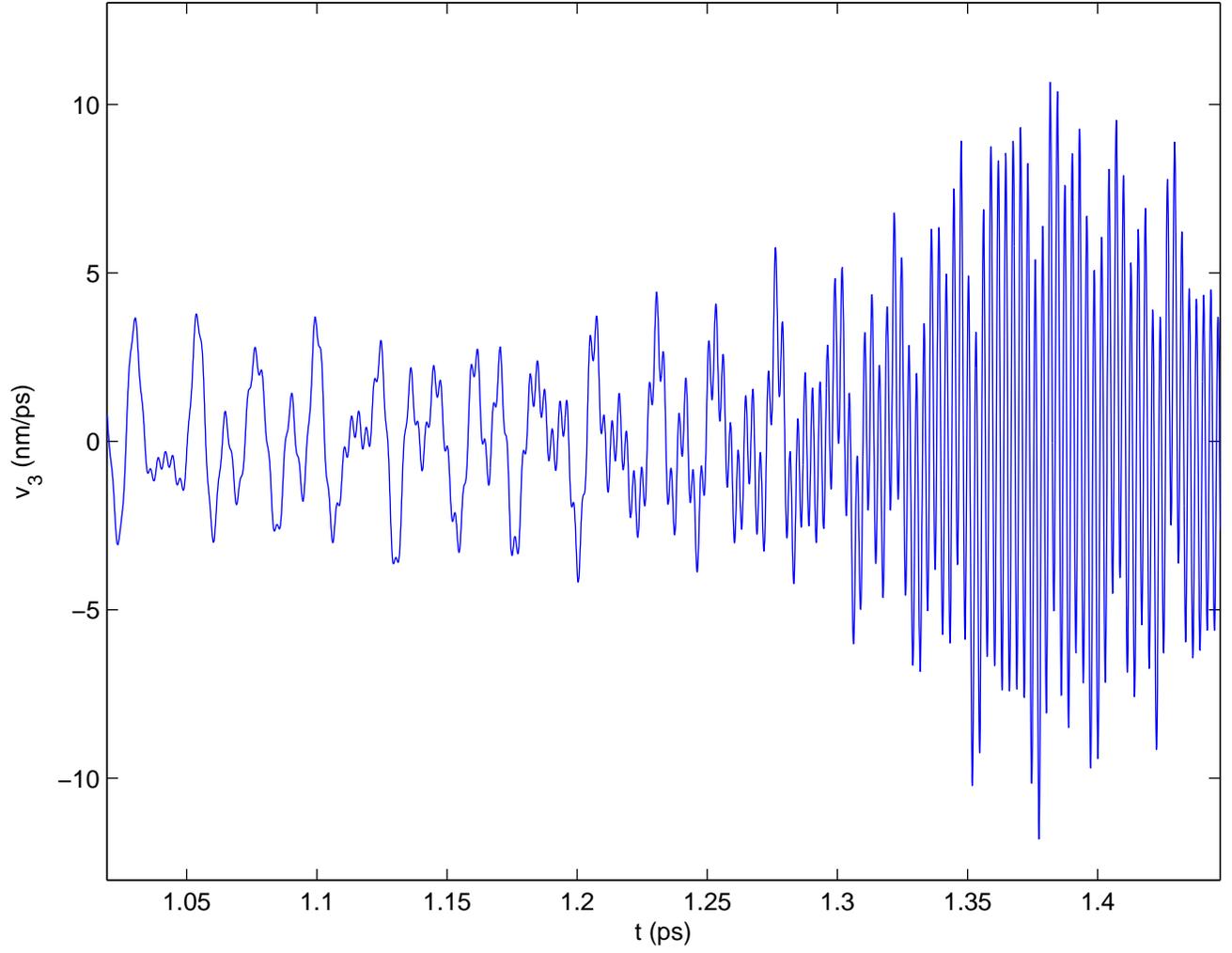}
\caption{The velocity of the wall between box 3 and box 4 around the time of the transparency
instability shows the slow, low amplitude oscillations giving way to the fast, large
amplitude oscillations driven by the growing fluctuating component of the Fermi pressure.}
\label{v3gamma0Mass100}
\end{figure}

We first describe the time evolution without the F\"orster term (i.e., $\beta_j=0$).
A typical time sequence is shown in Fig.~\ref{gamma0Mass100}.  Here the detuning parameter
$\Delta\mu=0.02$, so that the
effective electron masses were $1, 1.02, 1.04, 1.06$, and the wall masses were all $M=100$.
Contrary to intuition, the excitation placed initially in box 1 does not then move to box 2.  Rather,
after a delay, it suddenly appears in box 4.  Thereafter, since energy is conserved, and there is no
dissipation mechanism, it bounces back to box 3, back to box 2, and eventually becomes more chaotically
distributed, but tending to stay at the bottom of the funnel (box 4).  The reason for this
is that once the states evolve into superposition states (and this takes some time, around  1 ps in
Fig.~\ref{gamma0Mass100}), there is no barrier to energy flowing coherently through the complex, as the oscillating
Fermi pressure dominates the time evolution, inducing quantum transitions throughout
the system.  The change in the nature of the oscillations can be seen
in the velocity of the third wall, Fig.~\ref{v3gamma0Mass100}.  The slow oscillation of the wall about
its equilibrium, governed mainly
by the static steric pressure, gives way to rapid oscillations driven
by the fluctuating component of Fermi pressure once the fluctuating component
is large enough, at around $t=1.3$~ps.
The existence of a fluctuating component in one box drives Rabi transitions in the neighboring boxes
that increase their fluctuating
components, so that the process feeds on itself.  This is the reason for the instability.
It would be exponential growth if it were not bounded by unitarity.

\begin{figure}
\includegraphics{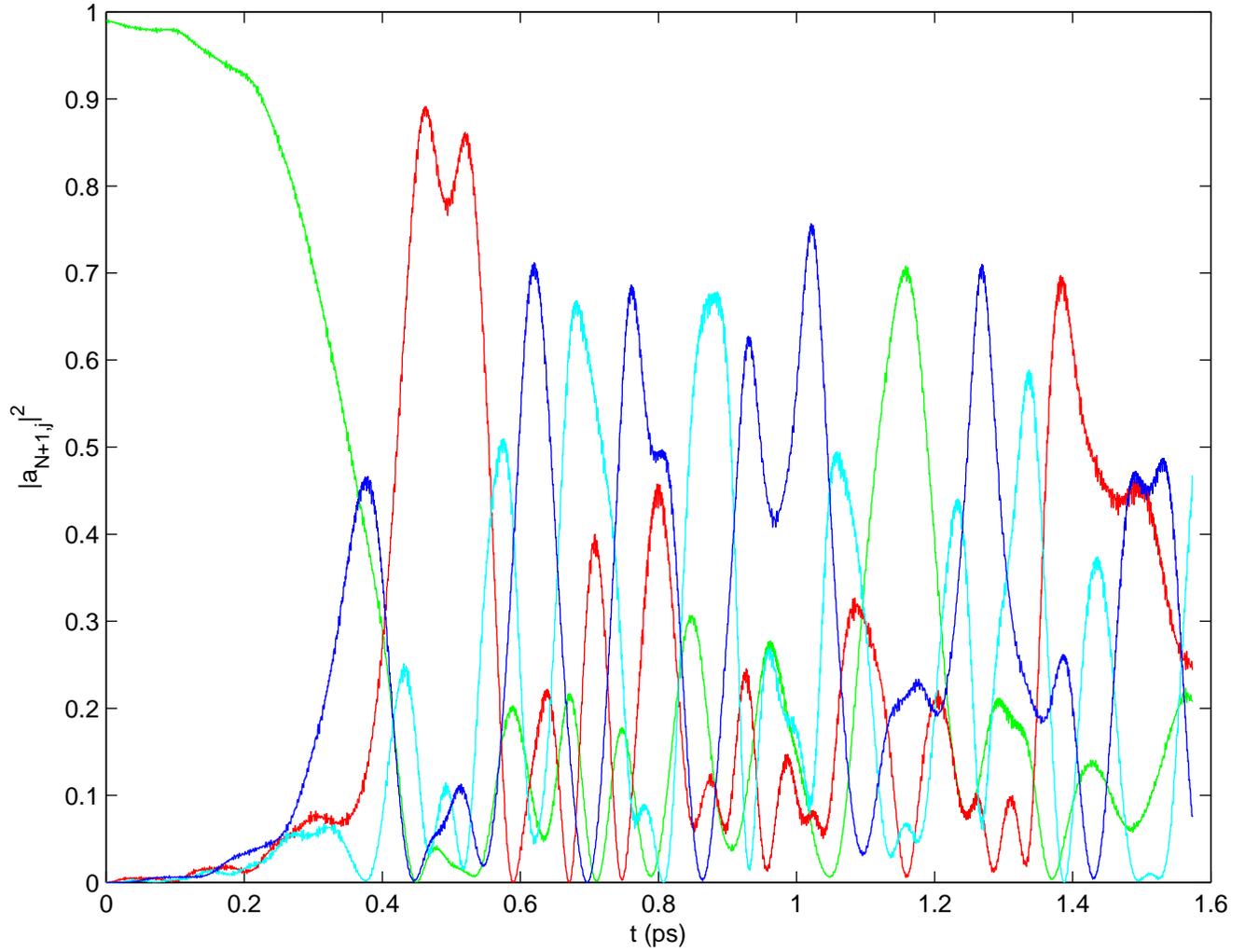}
\caption{The excitation $|a_{N+1}|^2$ as a function of time, as in Fig.~\ref{gamma0Mass100},
but with a slight admixture of the ground state in box 1, so that the Fermi
pressure fluctuations were already appreciable at t=0.
The excitation now reaches the last (4th) box directly from the first box,
with about 45\% strength, in less than 400 fs.  Transparency of the system is
clear as the energy moves rapidly thereafter from box to box.}
\label{gamma0Mass100primed}
\end{figure}

If we change the initial conditions so that
box 1 is initially in the state
$\sqrt{0.99}|N^*>+\sqrt{0.01}|N>$, i.e., we add a slight admixture of the box 1 ground state,
the sequence that required a 1 ps delay now happens immediately, as in Fig.~\ref{gamma0Mass100primed}.
This demonstrates the existence of the instability by starting the system a little further along
in its unstable time development.  In either case energy transfers,
when they occur, are on a time scale of about
a hundred femtoseconds.

For each value of the detuning parameter $\Delta\mu$ there is a range of wall masses $M$ (all chosen the
same, for simplicity) that allows this self-induced transparency.
Despite the apparent universality of this mechanism, it still requires parameters
to be chosen appropriately.  If $M$ is too large, neighboring boxes are
never brought into resonance by the oscillations that follow the initial photoabsorption.
If $M$ is too small, the classical oscillations are large and
neighboring boxes spend too little time in resonance to evolve into
the mixed states that drive the Fermi pressure mechanism.

\begin{figure}
\includegraphics{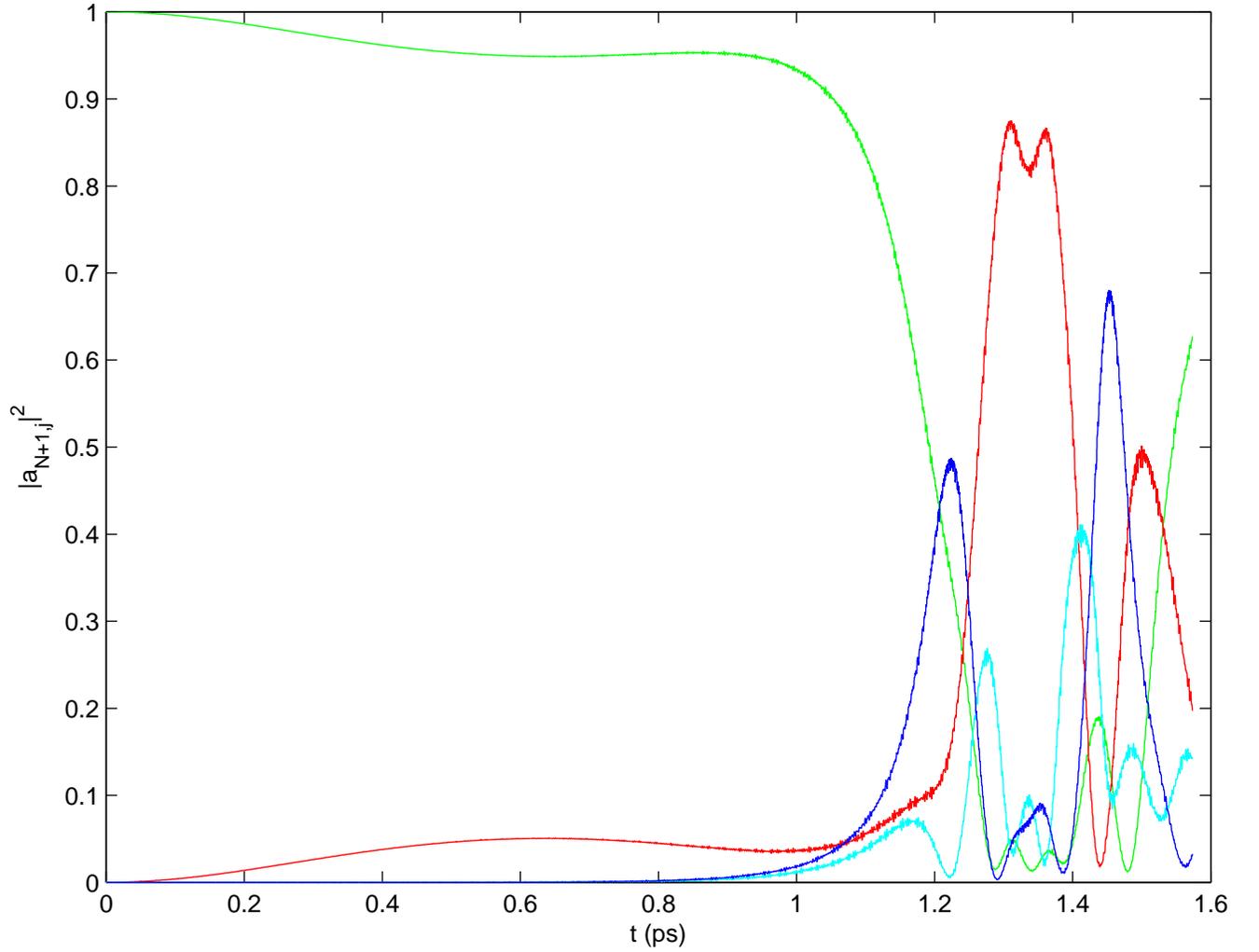}
\caption{The excitation $|a_{N+1}|^2$ as a function of time in box 1 (green), box 2 (red), box3 (cyan)
and box 4 (blue).  Here N=14, and the effective masses, creating the funnel,
were $\mu_1=1$, $\mu_2=1.02$, $\mu_3=1.04$, and $\mu_4=1.06$.  All wall masses were M=100.
 The F\"orster term is included with
a dimensionless coupling constant 0.04, as estimated in the text.}
\label{gamma04Mass100}
\end{figure}

\begin{figure}
\includegraphics{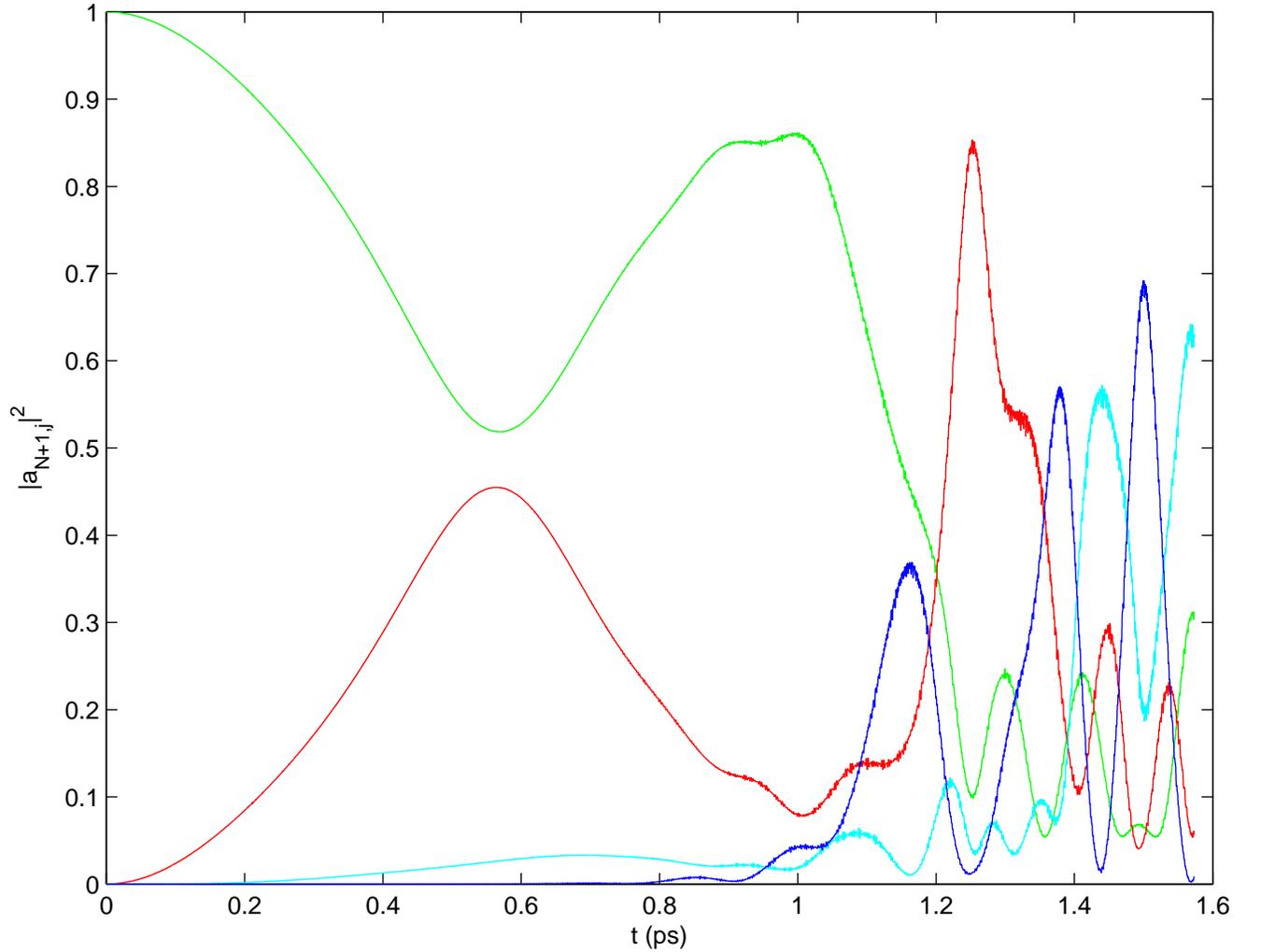}
\caption{The excitation $|a_{N+1}|^2$ as a function of time in box 1 (green), box 2 (red), box3 (cyan)
and box 4 (blue).  As before, N=14, and the effective masses, creating the funnel,
were $\mu_1=1$, $\mu_2=1.02$, $\mu_3=1.04$, and $\mu_4=1.06$.  All wall masses were M=100.
The F\"orster term is included with
a dimensionless coupling constant 0.1, corresponding to a F\"orster transition
time comparable to the time for Fermi pressure oscillations to develop.  The initial
transfer is now mainly to box 2, not box 4.}
\label{gamma1Mass100}
\end{figure}

Now we add the F\"orster term to the coherent evolution.  Using the coupling constant estimated in Section \ref{op}
we find the time evolution
of Fig.~\ref{gamma04Mass100}, only slightly different from Fig.~\ref{gamma0Mass100},
as one might expect, given the very different characteristic times for the two transfer processes that we
are considering.  With a stronger F\"orster coupling constant than we estimated above
(0.1 instead of 0.04), the transfer is very noticeably affected, and the transfer is now mainly to the adjacent
box 2 and not to box 4, as seen in Fig.~\ref{gamma1Mass100}.
The two mechanisms are not exactly in competition.  The coherent F\"orster interaction, by
mixing excited states and ground states, actually speeds up the onset of the Fermi pressure process,
as one sees in the peak corresponding to box 4, now just beyond 1 ps.
\section{Discussion}
The Fermi pressure
mechanism that we have described does transfer energy down a detuning gradient.
The sudden transfer of energy from the first to the last box
in the model is a counterintuitive and surprising feature.
It is due to a self-induced transparency that is initiated by an
instability in the oscillating Fermi pressure amplitude,
leading to its rapid growth.
This phenomenon could conceivably offer a dramatic speedup in energy transfer
within close packed molecular complexes.


\end{document}